\documentclass[aps,twocolumn,floats,superscriptaddress,showpacs]{revtex4}
\usepackage{amsmath,bm}%
\usepackage{graphicx}%

\begin{document}
%\twocolumn [\hsize\textwidth\columnwidth\hsize
%           \csname @twocolumnfalse\endcsname
%\draft
%\preprint{SINR/HIC-9904}
\title{Poissonian reducibility and thermal scaling within the lattice gas model
and molecular dynamics model}
\author{Y. G. Ma }
%\thanks{Present Address: Cyclotron Institute, Texas A \& M
%University, College Station, Texas 77801}

%\email{}

%\address{

\affiliation{China Center of Advanced Science and Technology
(World Laboratory),
 P. O. Box 8730, Beijing 100080, CHINA}
\affiliation{Shanghai Institute of Nuclear Research, Chinese
Academy of Sciences, P.O. Box 800-204, Shanghai 201800, CHINA}
%\footnotemark \footnotetext{Mailing address.}
\affiliation{LPC,
IN2P3-CNRS, ISMRA et Universit\'e, Boulevard Mar\'echal Juin,
14050 Caen Cedex, FRANCE }

\date{\today}
\begin{abstract}
The emission of clusters in the nuclear disassembly is
investigated within the framework of isospin dependent lattice gas
model and classical molecular dynamics model. As observed in the
recent experimental data, it is  found that the emission of
individual cluster is poissonian and thermal scaling is observed
in the linear Arrhenius plots made from the average multiplicity
of each cluster. The mass, isotope and charge dependent "emission
barriers" are extracted from the slopes of the Arrhenius plots and
their possible physical implications are investigated.
\end{abstract}
%\pacs{PACS Number(s):
  %    {25.70.Pq}{ Multifragment emission and correlations};
  %    {05.70.Jk}{ Critical point phenomena};
  %    {24.10.Pa}{ Thermal and statistical models};
  %    %{02.70.Ns}{Molecular dynamics and particle methods}
 %     }
%************************************************************************
%\vskip2pc] \vspace{1.cm}

\maketitle

%\narrowtext

%\section{Introduction}

In low-intermediate energy heavy ion collisions (HIC) the hot
nuclei with moderate temperature can be formed and they finally
deexcite by the different decay modes, such as the emission of
multiple intermediate mass fragment ($IMF$), $i.e.$
multifragmentation. Despite of the extensive studies in
experiments and theories, it is still difficult to clarify whether
the multifragmentation is statistical or dynamical, sequential or
simultaneous. Recently Moretto et al. found that there exists the
resilient reducibility and thermal scaling in multiple fragment
emission process, which gives one a helpful and clear picture to
look and understand the multifragmentation. They observed that the
experimental Z-integrated fragment multiplicity distributions
$P_n^m$ are binomially distributed,
\begin{equation}
P_n^m(p) = \frac{m!}{n!(m-n)!}p^n(1-p)^{m-n}
\end{equation}
 in each transverse energy ($E_t$) window, where $n$ is the number of emitted
fragments and $m$ is interpreted as the number of times the system
tries to emit a fragment. The probability of emitting $n$
fragments can be reduced to a single-particle emission probability
$p$ which gives linear Arrhenius plots ($i.e.$ excitation
functions) when $ln(1/p)$ is plotted vs 1/$\surd(E_t)$. By
assuming a linear relationship between $\surd(E_t)$ and
temperature $T$, the linearity of the observed $ln(1/p)$ vs
1/$\surd(E_t)$ plot can be explained to a thermal scaling of the
multifragment process \cite{More,Tso,More97}. In this case, these
linear Arrhenius plots suggest that $p$ has the Boltzman form $p
\propto exp(-B/T)$ with  a common fragment barrier $B$. However,
since the binomial decomposition has been performed on the
$Z$-integrated multiplicities, typically associated with $3 \leq Z
\leq 20$,  the Arrhenius plot generated with the resulting one
fragment probability $p$ is an average over a range of $Z$ values.
 More recently, Beaulieu, Phair and  Moretto et al.
 found that the fit with the binomial distribution can be also replaced
with the Poisson distribution in the constraint of charge
conservation.

Instead of analyzing for $Z$-integrated multiplicities, they
analyzed the behavior of individual fragment species of a given
$Z$ for higher resolution experimental data and noticed that the
$n$-fragment multiplicities $P(n)$ obey a nearly Poisson
distribution,
\begin{equation}
P(n) = \frac{<n>^n e^{-<n>}}{n!},
\end{equation}
where $n$ is the number of fragments of a given Z and the average
value $<n>$  is a function of the total transverse energy $E_t$,
and were thus reducible to a single-fragment probability
proportional to the average value $<n>$ for each $Z$ \cite{Beau}.
Similarly the $<n>$ is found to be proportional to $exp(-B/T)$
providing that $T \propto \surd E_t$, $i.e.$ there exists also a
thermal scaling law. As pointed out by Moretto et al.
\cite{More99}, this kind of "reducibility" and "thermal scaling"
are empirically pervasive features of nuclear multifragmentation.
"Reducibility" proves nearly stochastic emission process. "Thermal
scaling" gives an indication of thermalization. More recently,
Elliott and Moretto et al. discovered that the common features of
Poissonian reducibility and thermal scaling can also be revealed
in percolation and the Fisher droplet model \cite{Elliott}. Of
course, we should keep in mind that the assumption of $\surd(E_t)
\propto T$ may be only valid for compound nuclei formed at
low-to-moderate temperatures, but fail at higher temperatures in
the experimental data \cite{Tsang}. In the present work, we will
adopt the true temperature to study the reducibility and thermal
scaling even though  the assumption of $ \surd(E_t) \propto T$ is
found to be also valid at low-to-moderate temperatures in the
present model calculation. By investigating the mean cluster
multiplicity as a function of temperature, we will illustrate that
the Poissonian reducibility and thermal scaling is also valid for
fragment emission in nuclear disassembly via the theoretical
reexamination of thermal equilibrium models.

In this Letter, we will analyze the fragment multiplicity distributions for
each individual fragment $Z$ and $A$ value in the framework of isospin
dependent lattice gas model (I-LGM) and classical molecular dynamics (I-CMD).
We will show that they are Poissonian and the associated mean multiplicities
for each $Z$ or $A$ give linear Arrhenius plots as the experimental data
illustrated in \cite{Beau}. The $A$ and $Z$  dependent barriers are
extracted and investigated as a function of source size. Within our
knowledge, this is the first time to explore the Poissonian reducibility
and its thermal scaling for the individual fragment in the nuclear disassembly
within the lattice gas model and molecular dynamics model.

%\section{Models}

 The lattice gas model was developed to describe
the liquid-gas phase transition for atomic system by  Lee and Yang \cite{Yang52}.
The same model has already been applied to nuclear physics for isospin
symmetrical systems in the grand canonical ensemble \cite{Biro86} with a sampling
of the canonical ensemble
\cite{Jpan95,Jpan96,Mull97,Camp97,Jpan98,Gulm98,Carmona98,Ma99},
and also for isospin
asymmetrical nuclear matter in the mean field approximation \cite{Sray97}.
In addition, a classical molecular dynamical model is used to compare with
the results of lattice gas model. Here we will make a brief description
for the models.

In the lattice gas  model, $A$ (= $N + Z$) nucleons with an occupation number
$s_i$ which is defined $s_i$ = 1 (-1) for a proton (neutron) or $s_i$ = 0 for
a vacancy, are placed on the $L$ sites of lattice. Nucleons in the nearest
neighboring sites have interaction with an energy $\epsilon_{s_i s_j}$.
The hamiltonian is written as
\begin{equation}
E = \sum_{i=1}^{A} \frac{P_i^2}{2m} - \sum_{i < j} \epsilon_{s_i s_j}s_i s_j ,
\end{equation}
where $P_i$ is the momentum of the nucleon and $m$ is its mass.
The interaction constant $\epsilon_{s_i s_j}$ is chosen to be isospin dependent
and be fixed to reproduce the binding energy of the nuclei \cite{Jpan98}:
\begin{eqnarray}
 \epsilon_{nn} \ &=&\ \epsilon_{pp} \ = \ 0. MeV \nonumber , \\
 \epsilon_{pn} \ &=&\ - 5.33 MeV.
\end{eqnarray}
 Three-dimension cubic lattice with $L$ sites is used which
results in $\rho_f$ = $\frac{A}{L} \rho_0$ of an assumed freeze-out density
of disassembling system, in which $\rho_0$ is the normal nuclear density.
The disassembly of the system is to be calculated at $\rho_f$, beyond
which nucleons are too far apart to interact.  Nucleons are put into
lattice by Monte Carlo Metropolis sampling. Once the nucleons
have been placed we also ascribe to each of them a momentum by Monte Carlo
samplings of Maxwell-Boltzmann distribution.

Once this is done the I-LGM immediately gives the cluster distribution
using the rule that two nucleons are part of the same cluster if
\begin{equation}
 P_r^2/2\mu - \epsilon_{s_i s_j}s_i s_j < 0 ,
\end{equation}
where $P_r$ is the relative momentum of two nucleons and $\mu$ is their
reduced mass. This prescription is evidenced to be similar to the Coniglio-Klein's
prescription \cite{Coni80} in condensed matter physics and be valid in I-LGM
\cite{Camp97,Jpan96,Jpan95,Gulm98}. To calculate clusters using I-CMD we
propagate the particles from the initial configuration for a long time under
the influence of the chosen force. The form of the force is chosen to be also
isospin dependent in order to compare with the results of I-LGM. The potential
for unlike nucleons is
\begin{eqnarray}
 v_{\rm n p}(r) (\frac{r}{r_0}<a)\ &=&\ A\left[B(\frac{r_0}{r})^p-(\frac{r_0}{r})^q\right]\nonumber
    exp({\frac{1}{\frac{r}{r_0}-a}}), \\
v_{\rm  n p}(r) (\frac{r}{r_0}>a)\ &=&\ 0.
\label{pot}
\end{eqnarray}
In the above, $r_0 = 1.842 fm$ is the distance between the centers of two adjacent
 cubes. The parameters of the potentials are $p$ = 2, $q$ = 1, $a$ = 1.3,
$B$ = 0.924, and $A$ = 1966 MeV. With these parameters the
potential is minimum at $r_0$ with the value -5.33 MeV, is zero
when the nucleons are more than 1.3$r_0$ apart and becomes
stronger repulsive when $r$ is significantly less than $r_0$. The
potential for like nucleons is written as
\begin{eqnarray}
v_{\rm p p}(r) ( r < r_0 )\ &=&\  v_{\rm n p}(r)- v_{\rm n p}(r_0)\nonumber , \\
v_{\rm p p}(r) ( r > r_0 )\ &=&\ 0.
\end{eqnarray}
This means there is a repulsive core which goes to zero at $r_0$ and is zero
afterwards. It is consistent with the fact that we do not put two like
nucleons in the same cube.
The system evolves for a long time from the initial configuration obtained
by the lattice gas model under the influence of  the above potential.
At asymptotic times the clusters are
easily recognized. The cluster distribution and the quantities based on it in
the two models can now be compared. In the case of proton-proton interactions,
the Coulomb interaction can also be added separately and compared with the
cases where  the Coulomb effects are ignored.

%\section{Results and Discussions}

In this Letter we  choose the medium size nuclei $^{129}$Xe  as a
main example to analyze the behavior of individual fragment
emission during nuclear disassembly with the helps of I-LGM and
I-CMD. In addition, the systems with $A_{sys}$ = 80 ($Z_{sys}$ =
33) and 274 ($Z_{sys}$ = 114) are also studied to investigate the
possible source size dependence. In  most case, $\rho_f$ is chosen
to be about 0.38 $\rho_0$, since the experimental data can be best
fitted by $\rho_f$ between 0.3$\rho_0$ and 0.4$\rho_0$ in the
previous LGM calculations \cite{Jpan95,Beau96}, which corresponds
to  $7^3$ cubic lattice is used for Xe, $6^3$ for $A_{sys}$ = 80
and $9^3$ for $A_{sys}$ = 274 . In the condition of the fixed
freeze-out density, the only input parameter of the models is the
temperature $T$. In the I-LGM case, $\rho_f$ can be thought as the
freeze-out density but in the I-CMD case $\rho_f$ is, strictly
speaking, not a freeze-out density but merely defines the starting
point for time evolution. However since classical evolution of a
many particle system is entirely deterministic, the initialization
does have in it all the information of the asymptotic cluster
distribution, we will continue to call $\rho_f$ as the freeze-out
density. 1000 events are simulated for each $T$ which ensures
enough statistics.

%\subsection{Poissonian Reducibility}

One of the basic characters of the Poisson distribution Eq.(2) is
the ratio $\sigma_{n_i}^2/<n_i> \rightarrow 1$ where
$\sigma_{n_i}^2$ is the variance  of the  distribution and $<n_i>$
is the mean multiplicity. The first step we will show is this
ratio. We give these ratios for clusters classified with different
masses ($A$), light isotopes ($ISO$) and atomic numbers ($Z$) for
the disassembly of $^{129}Xe$ as a function of temperature in the
framework of I-LGM and I-CMD with Coulomb in Figure 1. Obviously,
the ratios are close to one except for protons, which indicates
that they might belong to the Poisson distributions. This can be
further supported by the multiplicity distribution in different
temperature. For instance, Fig.2  show the quality of the Poisson
fits to the charged particle multiplicity distribution for
$^{129}Xe$ in the I-LGM case. These Poisson fits are excellent for
all $Z \geq 2$ over the entire range of $T$. The same good Poisson
fit is obtained in the cases of I-CMD. Thus we can conclude that
Poissonian reducibility is valid in the thermal-equilibrium
lattice gas model or molecular dynamics.
\begin{center}
\begin{figure}
\includegraphics[scale=0.40]{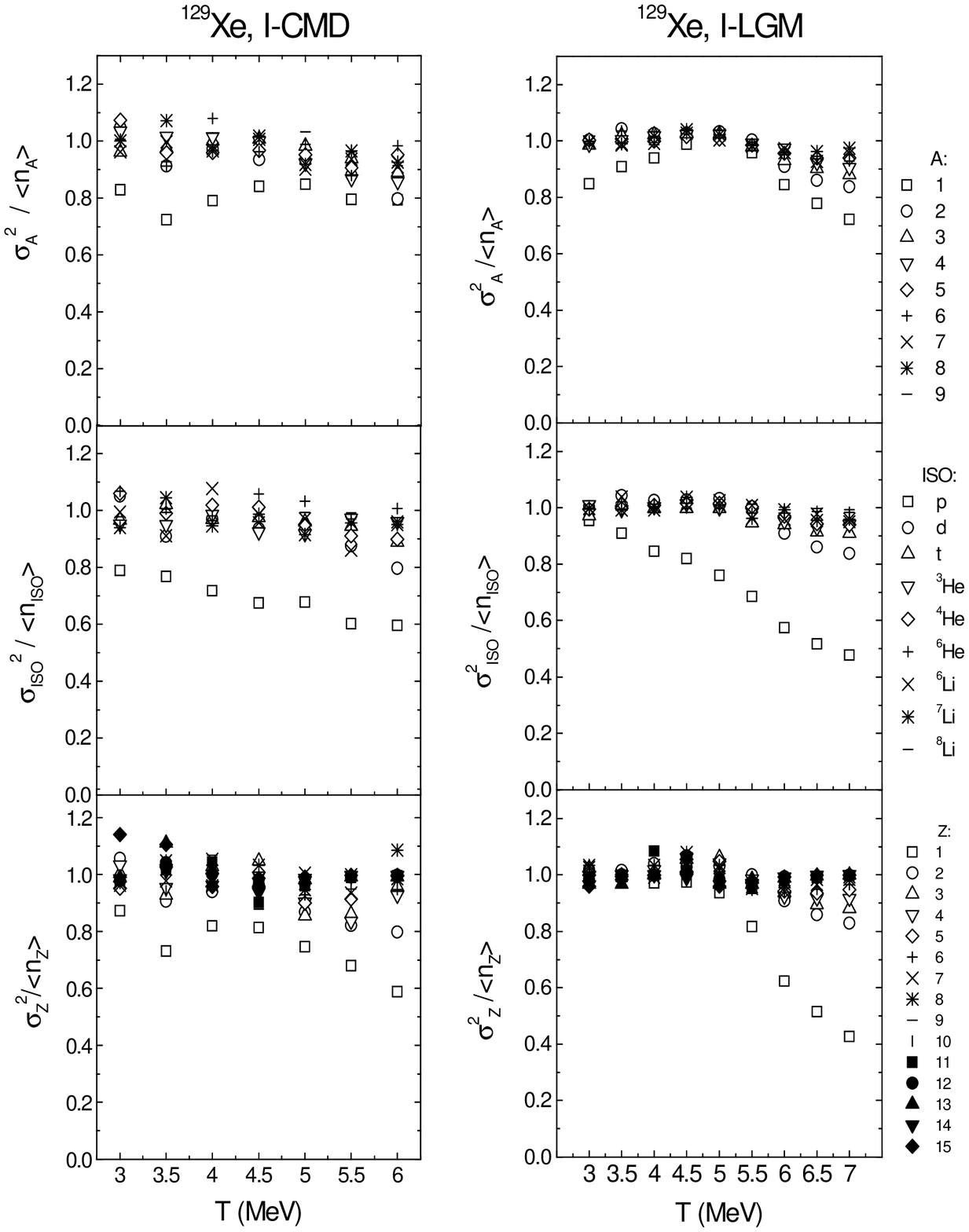}
\caption{\footnotesize The ratio of ${\sigma_i}^2/<n_i>$ for the
clusters classified with mass, light isotope mass and atomic
number as a function of temperature. The left panel is for the
I-LGM calculation and the right for I-CMD with Coulomb. The
symbols are illustrated on the figure.}
\end{figure}
\end{center}

%\subsection{Thermal Scaling}

To verify thermal scaling in the models, the temperature
dependence of the mean yield of clusters  is investigated.
Consequently, we generate Arrhenius plots by plotting $ln<n_Z>$ vs
$1/T$. Figure 3 give a family of these plots for the disassembly
of $^{129}Xe$ within the framework of I-LGM (left panel) and I-CMD
with Coulomb interaction (right panel).  The observed Arrhenius
plots are striking linear for the lower $T$ side, and their slopes
generally increase with increasing $Z$ value. Generally, the
thermal scaling is expected when the yields, for a fixed nucleon
number system, are dominated by fragment binding. This is  the
case when the temperatures are low compared to the binding energy
per particle. At these temperatures, one can anticipate one large
fragment surrounded by many small clusters. The contrary tendency
reveals in the high $T$ side where $ln<n_Z>$ increases with $1/T$,
$i.e.$ decreases with increasing $T$. In this case, nuclear
Arrhenius plots of $<n>$ with $1/T$ are not valid but the
Poissonian reducibility still remains (see Fig. 2). This behavior
of $<n>$ at higher $T$ is related to the branch of the fall of the
multiplicity of $IMF$ ($N_{IMF}$) with $T$ where  the
disassembling system is in vaporization \cite{Ogil,Tsang93,Ma95}
and hence only the lightest clusters are dominated and the heavier
clusters become fewer and fewer with increasing $T$. Afterwards we
will focus on the branch of lower temperature to discuss the
Arrhenius law. The overall linear trend illustrates that thermal
scaling is also present when the individual fragments of a
specific $A$, $ISO$ and  $Z$ are considered.
\begin{center}
\begin{figure}
\includegraphics[scale=0.40]{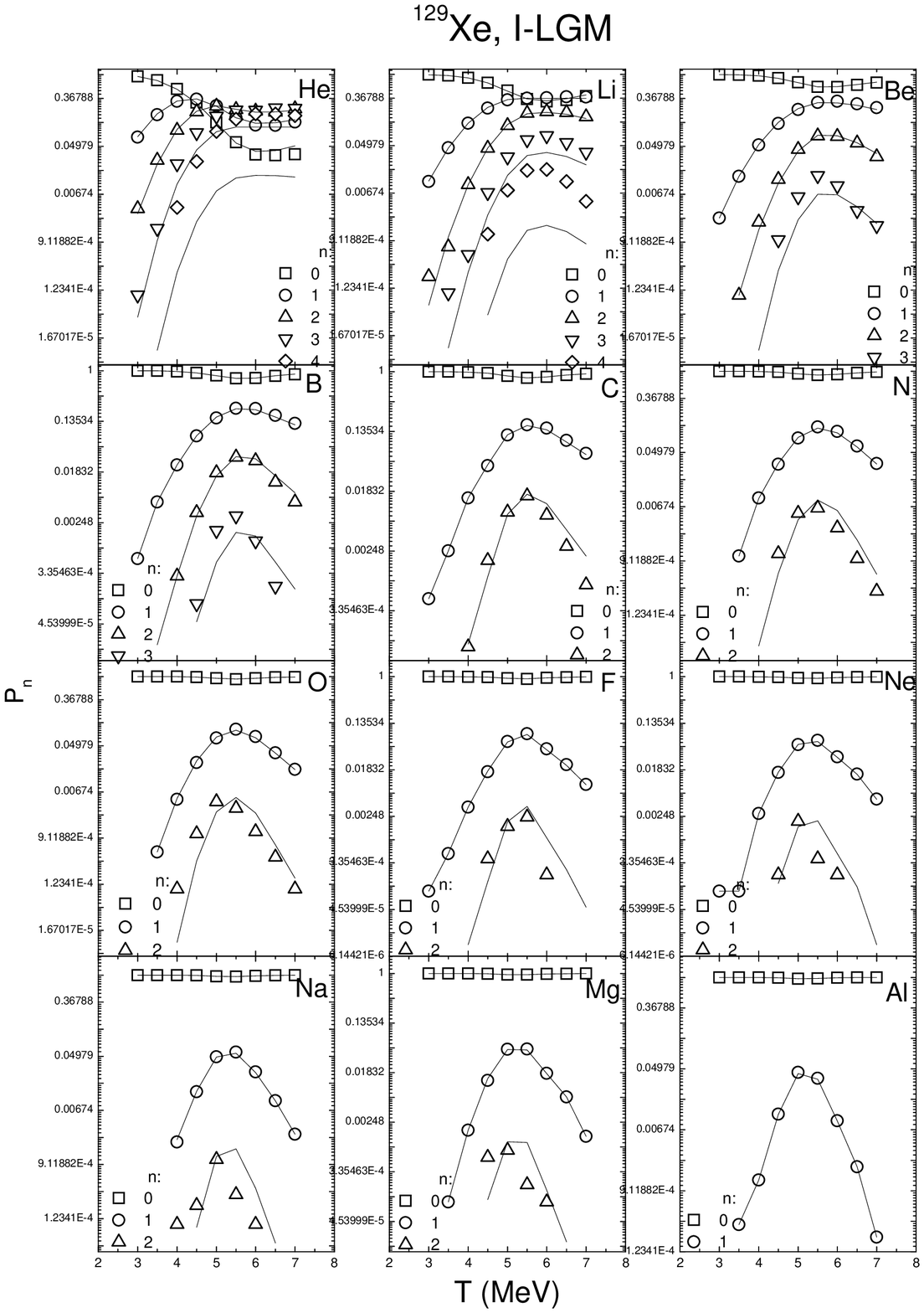}
\caption{\footnotesize The excitation functions $P_n$ for elements
with $Z \geq 2$ emission from the source $^{129}Xe$ in the I-LGM
calculation. The lines are Poisson fits with Eq.(2).}
\end{figure}
\end{center}

%\subsection{Fragment Dependent Barriers}

From figure 3 the slope parameter can be directly  extracted in
the lower $T$ side as a function of $Z$ or $A$. In Ref.
\cite{Beau} Moretto et al. has interpreted these slope parameters
as "emission barriers" of specific individual fragments. Figure 4
gives the emission barrier of individual fragments with different
$A$, $ISO$ and $Z$ in the framework of I-LGM, I-CMD with/without
Coulomb interaction. The error bar in the figure represents the
error in the extraction of the slope parameter.
 The first indication from this figure is that the emission
barrier in the I-LGM case is the nearly same as the  I-CMD case
without Coulomb force, which supports  that I-LGM is equivalent to
I-CMD without Coulomb interaction rather well when the nuclear
potential parameter is moderately chosen, but I-LGM is a quick
model to analyze the behavior of nuclear dissociations. The
inclusion of long-range Coulomb interaction makes the emission
barrier of individual fragments much lower since the repulsion of
Coulomb force reduces the attractive role of potential and hence
make clusters escape easily. The second indication is that the
emission barriers increase with $A$ ($Z$) at low $A$ ($Z$) values
and tend to be saturated at high $A$ ($Z$) ones. Similar
experimental results have been observed  for individual fragments
with different $Z$ in Ref.\cite{Beau} or different $A$ in
Ref.\cite{Elliott}  . However, the middle panel of Fig. 4 shows
that bare dependence of emission barrier of $ISO$ on $A$ in the
fixed atomic number Z, which indicates that the  Z dependence of
barrier is perhaps more intrinsical
 the A dependence is  mostly due to the
average effect over the species with the same A but different Z.

\begin{center}
\begin{figure}
\includegraphics[scale=0.40]{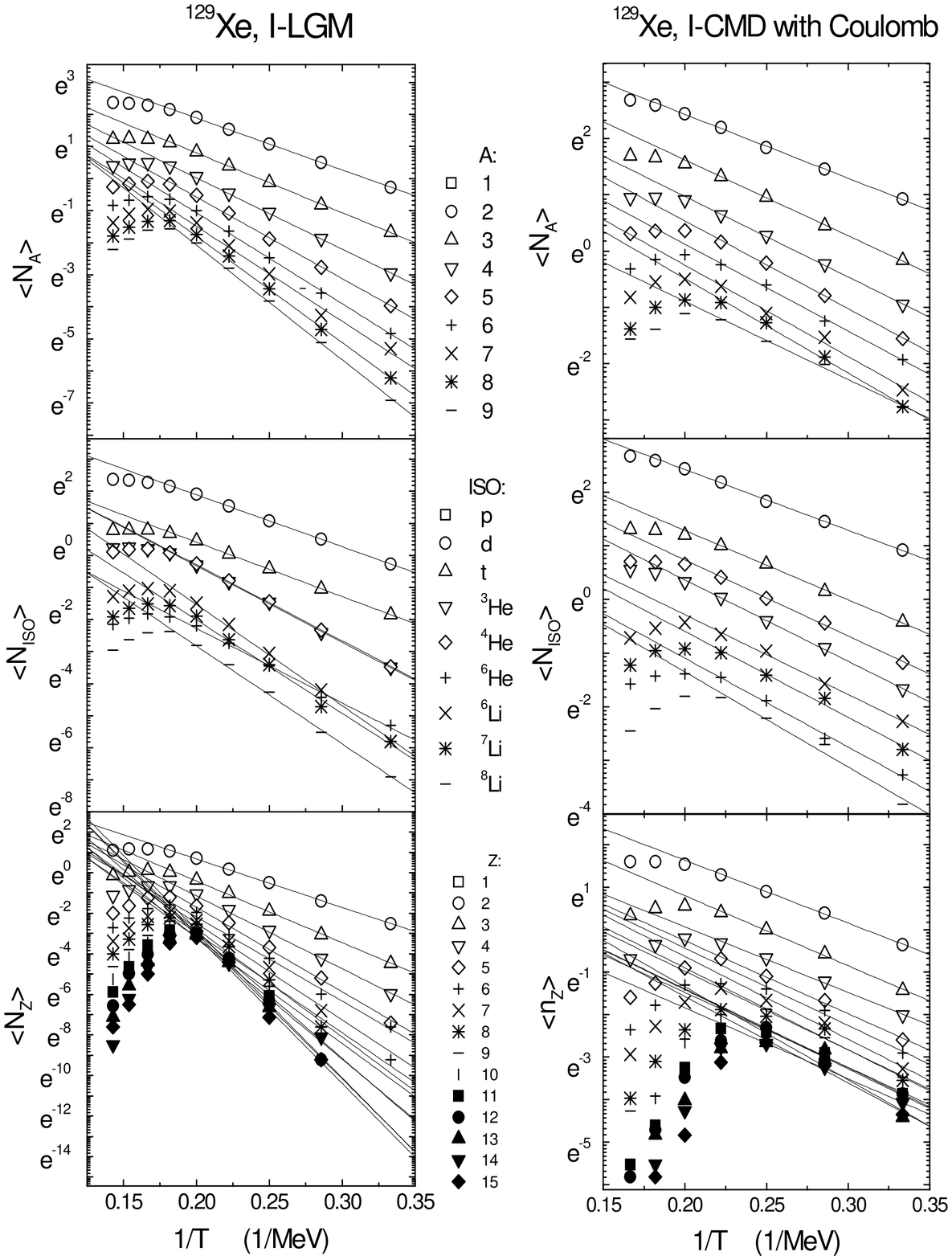}
\caption{\footnotesize The average yield per event of different
clusters classified with $A$ (top), $ISO$ (middle) and $Z$
(bottom) as a function of $1/T$. The left panel is for the I-LGM
calculation and the right for I-CMD with Coulomb. The solid lines
are fits to the calculations using a Boltzmann factor for $<n_i>$.
 The symbols are illustrated on the
figure.}
\end{figure}
\end{center}

%\subsubsection{Source size dependence}

On the origin of these barrier, the surface energy and Coulomb
energy would play the roles. If the cluster emission is mainly
controlled by its surface
 energy, it would suggest barriers proportional
to $Z^{2/3} (A^{2/3})$.  In the case of I-LGM and I-CMD
without Coulomb, we can try to fit the barrier for the
particles with different mass number by
\begin{equation}
B_{Coul. off} = c_1 \times {A_i}^{2/3}  ,
\end{equation}
or for the particles with different charge number by
\begin{equation}
B_{Coul. off} = c_1 \times ( (A/Z)_{fit} * Z_i) ^{2/3}  ,
\end{equation}
where $(A/Z)_{fit}$ is a fit coefficient of A/Z for emitted
particles, and $A_i$ ($Z_i$) is the mass (charge) of particle.
$c_1$ is the fit constant for surface energy term. The solid line
in the Fig.4a is a function of Eq.(8) with $c_1$ = 8.469 and the
solid line in the Fig.4c is a function of Eq.(9) with $c_1$ =
8.469 and $(A/Z)_{fit}$ = 1.866. These excellent fits imply that
the surface energy play a major role in controlling the cluster
emission when the long range Coulomb force is not considered.
However for the cluster emission with the Coulomb field, we can
assumed that the barrier  is mainly constituted by the surface
energy term and an additional Coulomb term as
%\begin{eqnarray}
\begin{widetext}
\begin{equation}
B_{Coul. on} = c_2 \times  A_i^{2/3} - %\nonumber \\
              \frac{1.44 \times A_i/(A/Z)_{fit} \times Z_{res}}
{r_{Coul} ({A_i}^{1/3} + ((A/Z)_{fit}*Z_{res})^{1/3})}  %\nonumber,\\
\end{equation}
\end{widetext}
%\end{eqnarray}
for the particles classified with different mass number, or
%\begin{eqnarray}
\begin{widetext}
\begin{equation}
B_{Coul. on}  =  c_2 \times  ( (A/Z)_{fit} * Z_i )^{2/3} - %\nonumber \\
  \frac{1.44 \times Z_i \times Z_{res}}
{r_{Coul} ( (Z_i*(A/Z)_{fit})^{1/3} + (Z_{res}*(A/Z)_{fit})^{1/3})}  %\nonumber,\\
\end{equation}
\end{widetext}
%\end{eqnarray}
for the particles  classified with different charge  number, where
$c_2$ is a fit constant for surface term and  $r_{Coul}$ is chosen
to be 1.22 fm. $Z_{res}$  is a fitted average charge number of the
residue. $(A/Z)_{fit}$ is chosen to be 1.866, as taken from the
fits for I-LGM. The overall fits for  $A$ and $Z$ dependent
barrier in the case of I-CMD with Coulomb force give $c_2$ =
12.921 and $Z_{res}$ $\sim$ 41 with the dot-dashed line in Fig.4a
and 4c. The excellent fit supports that
 the Coulomb energy plays another important role in the
cluster emission.

\begin{center}
\begin{figure}
\includegraphics[scale=0.40]{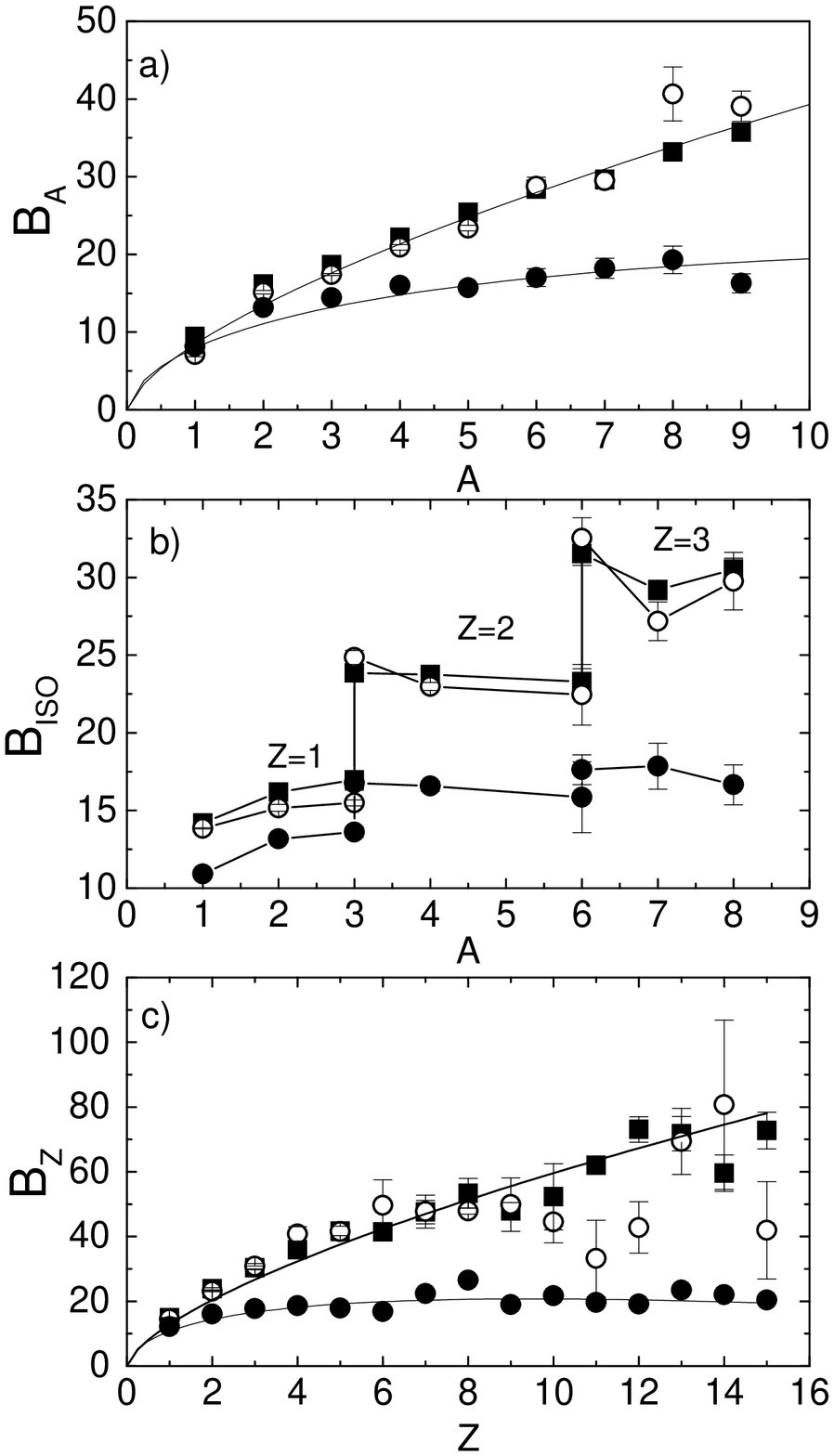}
\caption{\footnotesize The emission barrier extracted from the
Arrhenius plots as a function of cluster mass (top), isotopic mass
(middle) or cluster charge (bottom) in the cases of I-LGM (solid
squares), I-CMD without Coulomb (solid circles) and with Coulomb
(open circles). The solid lines are fits with the Eq. (8) or (9),
and the dot-dashed lines represent the fits with the Eq. (10) or
(11).}
\end{figure}
\end{center}

In the case of I-LGM and I-CMD without Coulomb, one would expect
the barrier for each $Z$ ($A$) to be nearly independent of the
system studied if only the surface energy is substantial to the
emission barrier. The left panel of the figure 5  shows the
results for $B_A$, $B_{ISO}$ and $B_Z$ for three different systems
in the I-LGM case. The same freeze-out density of 0.38$\rho_0$ and
the same $N/Z$ is chosen for the systems of $A_{sys}$ = 80 and
$A_{sys}$ = 274. Actually, it appears to have no obvious
dependence of emission barrier on source size as expected for the
role of surface energy. The solid line in the figure is the same
as in Fig.4.   However, when the long-range Coulomb interaction is
considered, the emission barrier reveals a source size dependence.
The right panel of figure 5 gives the emission barriers $B_A$,
$B_{ISO}$ and $B_Z$ in the case of I-CMD with Coulomb force. It
looks that the barrier increase with the decreasing of charge of
system, which can be explained with the Eq. (10) and (11) where
the decreasing of the residue $Z_{res}$ will result in the
decreasing of the Coulomb barrier and hence the increasing of the
emission barrier. The lines represent the fits with the Eq.(10)
and (11) for three different mass systems.
\begin{center}
\begin{figure}
\includegraphics[scale=0.40]{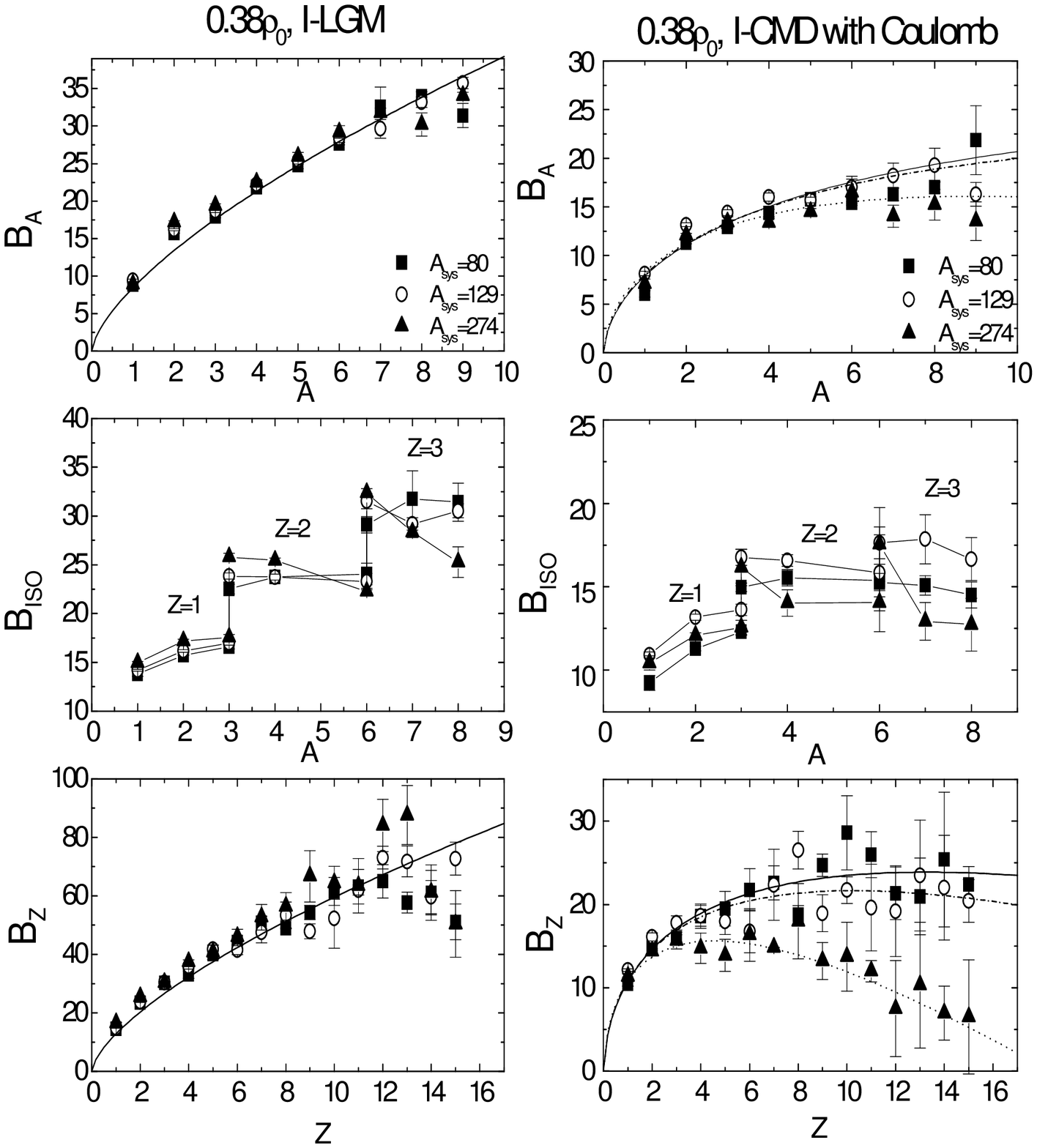}
\caption{\footnotesize The source size dependence of the  emission
barriers for the different clusters classified with mass (top),
isotopic mass (middle) or cluster charge (bottom) from in the
cases of I-LGM (left panel), I-CMD with Coulomb (right panel). The
lines in the left panel are fits with the Eq. (8) or (9), and the
solid, dot-dashed and dotted line in the right panel represents
the fits to the emission barrier of $A_{sys}$ = 80, 129 and 274,
respectively, with the Eq. (10) or (11).}
\end{figure}
\end{center}

In the above calculations, the freeze-out density of systems is
fixed at $\sim$ 0.38$\rho_0$.
 Considering the freeze-out density is an important debating variable in the latter stage of heavy ion
  collisions, here we will discuss the possible influence of freeze-out density on the emission barrier
   of clusters. The calculations at the freeze-out density  of 0.177$\rho_0$ and 0.597$\rho_0$ for $^{129}Xe$,
    corresponding to $9^3$ and $6^3$ cubic lattices respectively, are supplemented  to compare.
Figure 6 gives the results of $B_A$, $B_{ISO}$ and $B_Z$ at
different density. It looks that there are no obvious freeze-out
density dependence in the both cases of I-LGM and I-CMD. This is
also consistent  with that assumption  that the surface energy is
the dominant role in controlling the cluster emission.

\begin{center}
\begin{figure}
\includegraphics[scale=0.40]{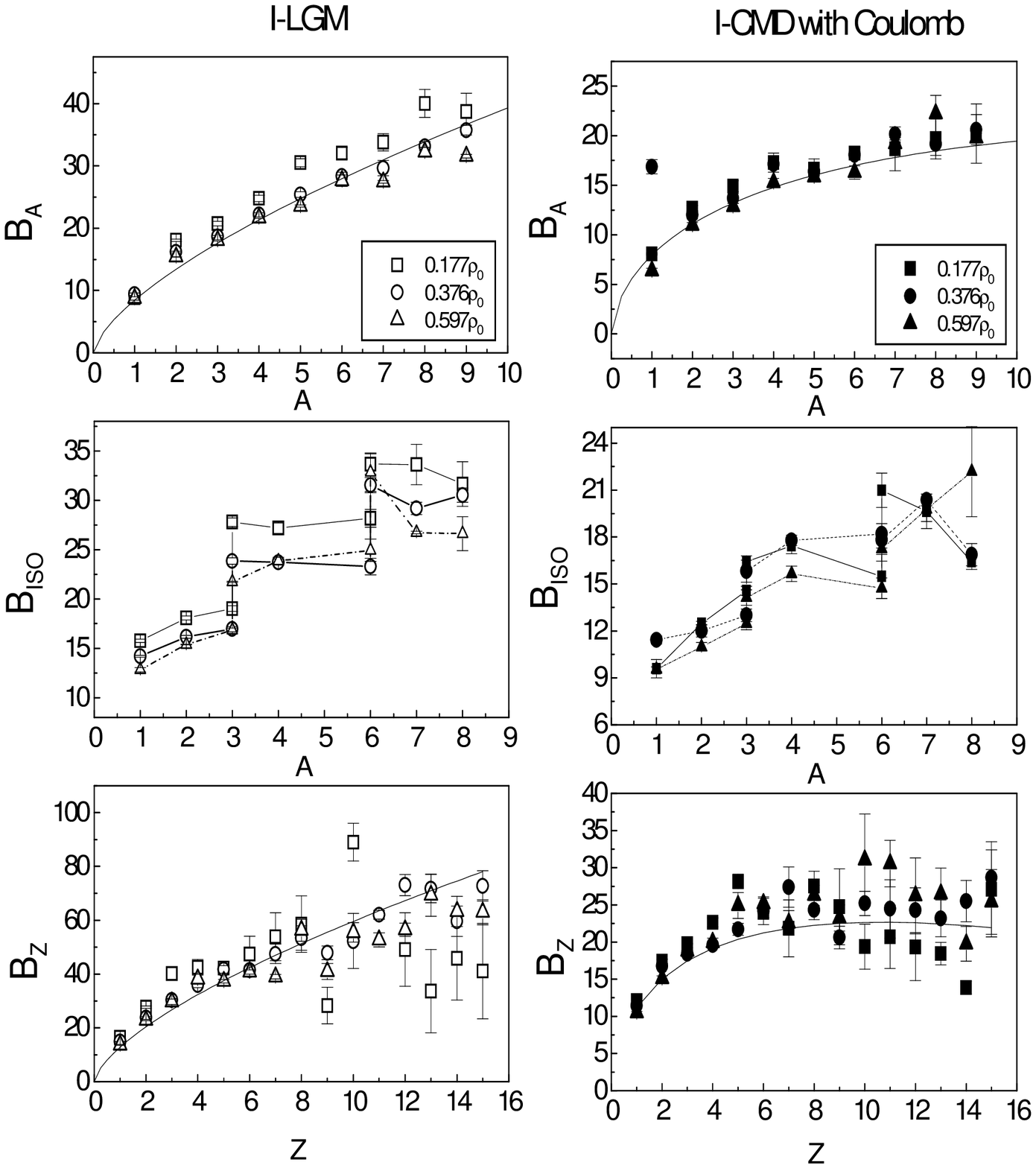}
\caption{\footnotesize The  emission barrier of $^{129}$Xe for the
different clusters classified with mass (top), isotopic mass
(middle) or cluster charge (bottom) at the different freeze-out
density in the cases of I-LGM (left panel), I-CMD with Coulomb
(right panel). The lines are fits with the Eq. (8) or (9) in the
left panel and
 with the Eq. (10) or (11 )  in the right panel.}
\end{figure}
\end{center}

In conclusion, the poisson reducibility and thermal scaling of the
emitted clusters is explored in the lattice gas model and
molecular dynamical model. The calculations are qualitatively
consistent with the recent experimental observation by
Moretto/Wozniak's group even though  the temperature is supposed
to be proportional to the total transverse energy in the latter
experiments. A systematic study of the emission barrier on the
cluster mass, isotope and charge proves that the cluster emission
is mainly controlled by both the surface energy and the Coulomb
interaction. In the framework of the lattice gas model and
molecular dynamics model without the Coulomb interaction, the
emission barrier relies on the cluster charge with the $Z^{2/3}$
($A^{2/3}$) law and it does not depend on the the source size and
freeze-out density, which indicates that the surface energy play a
basic dominant role to control the cluster emission. Conversely,
in the framework of molecular dynamics model with the Coulomb
force, the emission barrier will decrease strongly according to
the Eq.(10) and (11) and it decreases with the increasing of the
source size, illustrating that the Coulomb interaction also play
another weighty role to control the cluster emission.

\acknowledgments

Author would like to  thank Prof. B. Tamain, Prof. S. Das Gupta,
Prof W.Q. Shen and Dr. J.C. Pan for helps. This work was supported
in part by the NSFC for Distinguished Young Scholar under Grant
No. 19725521, the NSFC under Grant No. 19705012, the Special
Foundation of the President
 of Chinese Academy of Sciences, the Major State Basic Research Development
Program of China under Contract No. G200077400. It was also supported partly
by the IN2P3-CNRS Foundation of France.

%\begin{references}
{}
\end{document}